# Dephasing of an Electronic Two-Path Interferometer

I. Gurman[§], R. Sabo[§], M. Heiblum, V. Umansky, and D. Mahalu

*Braun Center for Submicron Research, Dept. of Condensed Matter physics,*

*Weizmann Institute of Science, Rehovot 76100, Israel*

*§ Both authors equally contributed to this work*

**This work was motivated by the quest for observing interference of fractionally charged quasi particles. Here, we study the behavior of an electronic Mach-Zehnder interferometer (MZI) at the integer quantum Hall effect (IQHE) regime at filling factors greater than one. Both the visibility and the drift velocity were measured, and found to be highly correlated as function of filling factor. As the filling factor approached unity, the visibility quenched, not to recover for filling factors smaller than unity. Alternatively, the velocity saturated around a minimal value at unity filling factor. We highlight the significant role interactions between the interfering edge and the bulk play, as well as that of the defining potential at the edge. Shot noise measurements suggest that phase-averaging (due to phase randomization), rather than single particle decoherence, is likely to be the cause of the dephasing in the fractional regime.**

Interference of multiple electron trajectories is generally used to better understand coherent electron phenomena and their dephasing processes. Yet, the implementation of a high visibility electronic interferometer faces challenges, such as restricting the number of trajectories and achieving sufficiently long coherence and thermal lengths. Therefore, chiral edge channels in the quantum Hall effect (QHE) regime [1], being 1D-like channels running along the edge of a two dimensional electron gas (2DEG), are ideal as distinct trajectories in electron interferometers [2, 3, 4]. Based on these edge channels, the electronic Mach-Zehnder interferometer (MZI) is a true two-path interferometer, with the enclosed magnetic flux between the two paths controlling the Aharonov-Bohm (AB) phase [5, 6]. Indeed, its practical success allowed innovative studies of coherence and dephasing [7, 8, 9], as well as studies of the nature of electron-electron interactions in the integer QHE regime [10, 11].

If realized in the fractional QHE regime, the MZI interference can serve as a probe of the statistics of fractionally charged quasiparticles [12, 13, 14, 15, 16]. However, despite a continuous effort to observe interference in the fractional regime, it was never observed. In an attempt to understand the absence of such interference, we mapped the linear and non-linear interference visibility as the filling factor approaches $v = 1$.

Experiments were conducted with a high mobility 2DEG, embedded in a GaAs-AlGaAs heterostructure with areal densities $n = (1.8 - 2.5) \cdot 10^{11} \text{cm}^{-2}$ and mobility of $(2.5 - 3.9) \cdot 10^6 \text{ cm}^2\text{V}^{-1}\text{s}^{-1}$ at electron temperature $(10 - 15)$ mK. The electronic MZI (see Fig. 1a) is defined by two quantum point contacts (QPCs), each acting as an electronic beam splitter, a modulation gate charged by $V_{mg}$, which controls the area $A$ pierced by the magnetic field $B$, and two drains – one inside the MZI (D2) and the other outside (D1). QPC1 splits the incoming electron beam into two paths, which interfere at QPC2, to be collected in by the two drains. Aharonov-Bohm interference oscillations depend on the enclosed magnetic flux $\Phi_{AB} = A \cdot B$ [17]. In addition, another QPC resides between the source and the MZI (not seen in the figure), and is used to separate the incoming edge channels such that different channels can arrive with different applied potential. Moreover, this QPC allows to separate the outer most edge channel from the conducting bulk when $R_{xx} \neq 0$. The non-linear transmission of the MZI is measured by applying DC+AC ($f_0 \approx 1$ MHz) to the source and measuring the AC component at D1. In turn, the transmission of the MZI obeys $t_{MZI} = t_1 t_2 + r_1 r_2 + 2\eta\sqrt{t_1 t_2 r_1 r_2}\cos(\Phi_{AB} + \varphi)$, where $t_i(r_i)$ is the transmission (reflection) probability of QPC$i$, $\varphi$ is an arbitrary (but constant) phase and $\eta$ is a coherence coefficient ($0 \leq \eta \leq 1$, depending on temperature, paths length and their difference, *etc*.). The visibility of the MZI is defined as $\frac{t_{max} - t_{min}}{t_{max} + t_{min}}$, where $t_{max(min)}$ is the maximal (minimal) phase dependent transmission of the MZI. Maximum visibility is obtained for $t_1 = 1 - t_2$, with the most convenient working point being $t_i = 0.5$, resulting in $t_{MZI} = \frac{1}{2} + \frac{\eta}{2}\cos(\Phi_{AB} + \varphi)$, with $\eta$ as the visibility.

With the outer most edge channel biased and the inner ones at ground potential, interference oscillations were measured in the linear ($V_{DC} = 0$) and non-linear ($|V_{DC}| > 0$) regimes. A typical linear transmission as function of $V_{mg}$, with visibility $\eta = 65\%$ and an average 0.5, is seen in the inset of Fig. 1b. Similar interference data was observed by varying the magnetic field instead of the area (not shown here). With $|V_{DC}| > 0$, the visibility exhibited the familiar lobe-like behavior as function of $V_{DC}$ (Fig. 1b), with distinct maxima and minima (a phase jump of $\pi$ in the oscillation appeared at each minima) [18]. The exact mechanism behind this effect is still debatable [19, 20, 21], but it is commonly accepted that electron-electron interactions are responsible for the 'self-dephasing' of the interferometer. Moreover, a single electron's wave-packet equals approximately the single path length $L$ at the first minima of the visibility $V_0$. This assumption allows extracting an approximate electron drift velocity in the interfering edge channel, $v_d = \frac{eV_0L}{h}$, where $e$ is the electron charge, $h$ is Plank's constant and $L$ is the single path length (in a symmetric MZI).

We start with a MZI with its edges defined by 'wet etching' (down to the donor layer; dubbed *etched-defined*), as shown in Fig. 1a. Interference was measured at different filling factors $\nu = hn/Be$, controlled by varying the magnetic field while keeping the areal electron density $n$ constant. The visibility and the electron velocity of two such similar devices were plotted as function of the filling factor in Fig. 1c. A clear correlation between the visibility (blue circles) and the electron velocity (green triangles) is evident [22]. The visibility reaches maximum around $\nu = 1.5$ and a minimum around $\nu = 2.5$; diminishing to zero as it approaches $\nu = 1$. The latter observation is surprising, since the outer edge channel is still well defined, with its conductance remaining within the $\sigma_{xy} = \frac{e^2}{h}$ plateau (bulk is still insulating). At the opposite end, $\nu > 3.5$, we find a finite tunneling current between the edge channels (likely, due to a smaller energy gap); therefore, edge channels are not well defined. Noting, as the visibility goes down from 80% to 20% (factor of four), the electron's drift velocity varies by a factor of three. However, while the visibility keeps on quenching, towards full dephasing around $\nu = 1$, the velocity saturates at $v_d \approx 1x10^6$ cm/s ($V_0 \approx 5\mu V$). The extracted velocities are an order or magnitude lower than the previously measured ones via edge magneto-plasmons excitation [23, 24, 25]. This discrepancy may be related to the inter-edge interactions (say, around $\nu = 2$), leading to slower neutral modes, which may dominate the periodicity of the 'lobe structure' [20, 26, 27].

Similar measurements were performed with the two paths confined by depleted regions under biased top gates (dubbed *gate-defined*; inset in Fig. 2a). In this MZI the confining potential is softer than the etched-defined one; yet, the dependence of the visibility on the bulk filling factor resembles (in general) the one in the etched-defined MZI. A noted difference is the random phase that accompanies the periodic AB oscillations, with an increasing rate as the filling factor approaches $\nu \approx 1$ (Fig. 2a - IV, V). This is as opposed to the $\nu > 1.5$ case, where the visibility decreases in a smooth manner (Fig. 2a – II, III). As in the etched-defined MZI, no sign of interference was found at $\nu < 1$.

In order to understand the reason for the different behavior of the two types of MZIs, the internal structure of the edge channels was studied in more detail. A 'two-QPC' configuration, which allows probing the conductance of the outer most edge channel, was utilized [28, 29]. For all integer states the conductance of the outer most edge was found to be $\frac{e^2}{h}$; namely, an integer channel of the lowest spin-split Landau level. However, at bulk filling $\nu = 1$ the conductance of the outer most edge channel was found to be $\frac{2}{3}\frac{e^2}{h}$ in the gate-defined configuration (Fig. 2b, red squares). Note that the appearance of the phase noise coincides with the formation of the 2/3 channel; with likely a counter-propagating neutral mode. Contrary, in the etched-defined case, such fractional state was not detected at the edge (Fig. 2b, black circles) yet it could have been merged with the inner channel and thus not observed [29].

Does the characteristic behavior of the MZI depend only on the filling factor? To address this question, a gate-defined MZI with a top gate (biased $V_{tg}$) was fabricated, allowing control of the electron density, $0 < n_{tg} < n_0 = 1.8 \cdot 10^{11}$ cm$^{-2}$) in the MZI (Fig. 3a). The visibility was found to depend solely on the local filling factor in the MZI [22]. Here too, an increasing phase noise appears once entering the $\nu < 1.5$ regime.

While all the data presented thus far was of the outer most edge channel (the lowest spin-split LL), it is interesting (and challenging) to check its universality by interfering the next inner edge channel at $\nu \geq 2$. In Fig. 3b we show the visibility of the outer most edge channel (blue) and the next inner one (red). The two dependencies almost overlap. Here too, once crossing the $\nu = 2$ filling (thus entering fractional states of the second spin-split LL) the visibility of the interfering inner edge channel quenches with an escalating phase noise. Note that a high visibility interference of an inner edge is not typical and only seldom observed.

What is the reason for the disappearance of the interference oscillation at $\nu \leq 1$ and near $\nu = 2.5$? Does every electron lose coherence (hence, $\eta = 0$), or maybe, each electron acquires a random phase and thus phase-averaging destroys the observed interference (namely, $\langle \cos(\Phi_{AB} + \varphi) \rangle = 0$)? Understanding of the relevant dephasing mechanism might be crucial for a possible recovering of the fractional interference. Alas, null visibility cannot distinguish between these two mechanisms. Yang *et al.* argued [6] that *shot noise* measurements can be utilized to distinguish between these two dephasing mechanisms. Recalling the electronic excess noise density $\Delta S = 2eI\, t_{MZI}(1 - t_{MZI})\alpha(T)$, with $I$ the impinging current and $\alpha(T)$ a reduction factor due to finite temperature [30, 31]; tuning $t_1 = 0.5$ and varying $t_2$, the device's transmission is $t_{MZI} = \frac{1}{2} + \eta\sqrt{t_2(1-t_2)}\cos(\Phi_{AB} + \varphi)$. The expected excess noise is then $\Delta S = 2eI\alpha(T)\left[\frac{1}{4} - \eta^2 t_2(1-t_2)\cos^2(\Phi_{AB} + \varphi)\right]$; hence, strongly dependent on $\eta$. If phase averaging is dominant, $\langle \cos^2(\Phi_{AB} + \varphi) \rangle = 0.5$, the excess noise will have a parabolic function of $t_2$. Alternatively, if single electron decoherence is dominant ($\eta = 0$), the excess noise will be independent of $t_2$. The considerable contrast between the two scenarios exemplifies the potency of noise measurement with a dephased MZI.

Performing noise measurements at $\nu < 1$ and with $t_1 = 0.5$, the normalized excess noise exhibits a clear $t_2$ dependence (blue circles) - though not 'parabolic'. Alternatively, around $\nu = 2.5$ (red squares) the noise is 'parabolic' with $\eta = 0.95$. Evidently, phase averaging plays a major role in the dephasing of the interferometer. It is worth mentioning that in all the cases when interference vanished (due to high temperature [6], with an asymmetric MZI, with an Ohmic contact inserted in one path [32] and at various fractional quantum Hall states), we had found a 'parabolic' noise dependence on $t_2$ with $\eta > 0.7$ - but never observed $\eta = 0$.

While the presented results do not single out the reasons for the lack of interference in the fractional quantum Hall regime, a few important observations surfaced: (*i*) The visibility and the chiral velocity trend similarly as function of the filling factor, suggesting that a lower velocity is accommodated with a stronger dephasing. A longer dwell time in the MZI can account for the drop in the visibility. (*ii*) As the lower velocity is associated with weaker confining electric field at the edge, phase averaging, due to a 'spatially wider edge channel', is more likely; (*iii*) The absence of interference at filling one and lower may suggests that lack of screening of bulk charge fluctuations may have a dramatic effect on the interference. Indeed charge fluctuations are observed (via phase noise) as the filling approaches unity.(*iv*) A fractional state ($\nu = 2/3$) is formed at the edge as the filling is close to one, simultaneously with the appearing of phase noise (clearly observed in the gate-defined device). Such hole-conjugate states were shown to be accompanied by a counter-propagating neutral mode(s), which may cause energy relaxation and dephasing. (*v*) The correlation between the visibility and the filling factor points at the significance the bulk has on the edge current [33, 34]. (*vi*) Loss of interference, in particular in the fractional regime, likely results from phase averaging, while each quasiparticle's state maintains its own coherence. Not being a fundamental incoherent process of the fractional excitations, there is a ray of hope in observing interference by minimizing the averaging effects.

M.H. acknowledges the partial support of the Minerva foundation, the German Israeli Foundation (GIF), and the European Research Council under the European Community's Seventh Framework Program (FP7/2007-2013)/ERC Grant agreement No. 339070. I.G. is grateful to the Azrieli Foundation for the award of an Azrieli Fellowship.

Figures

# FIGURE 1

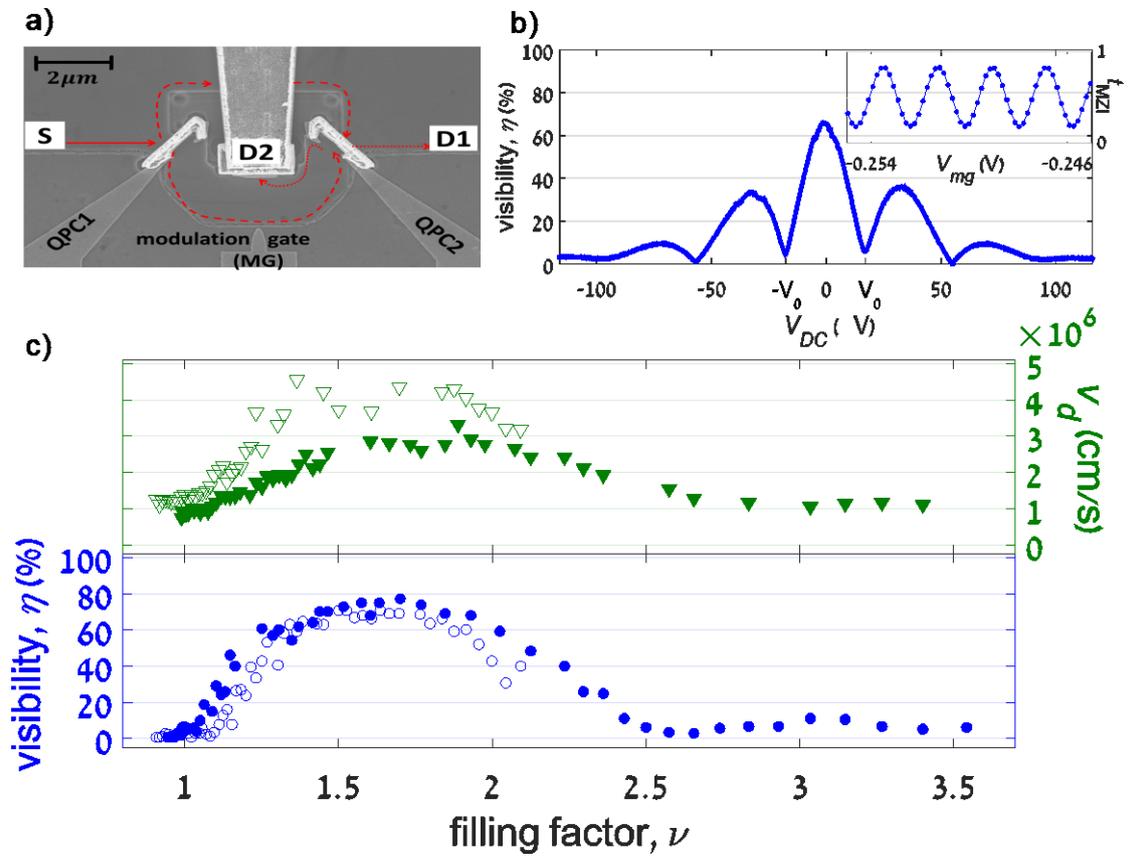

# FIGURE 2

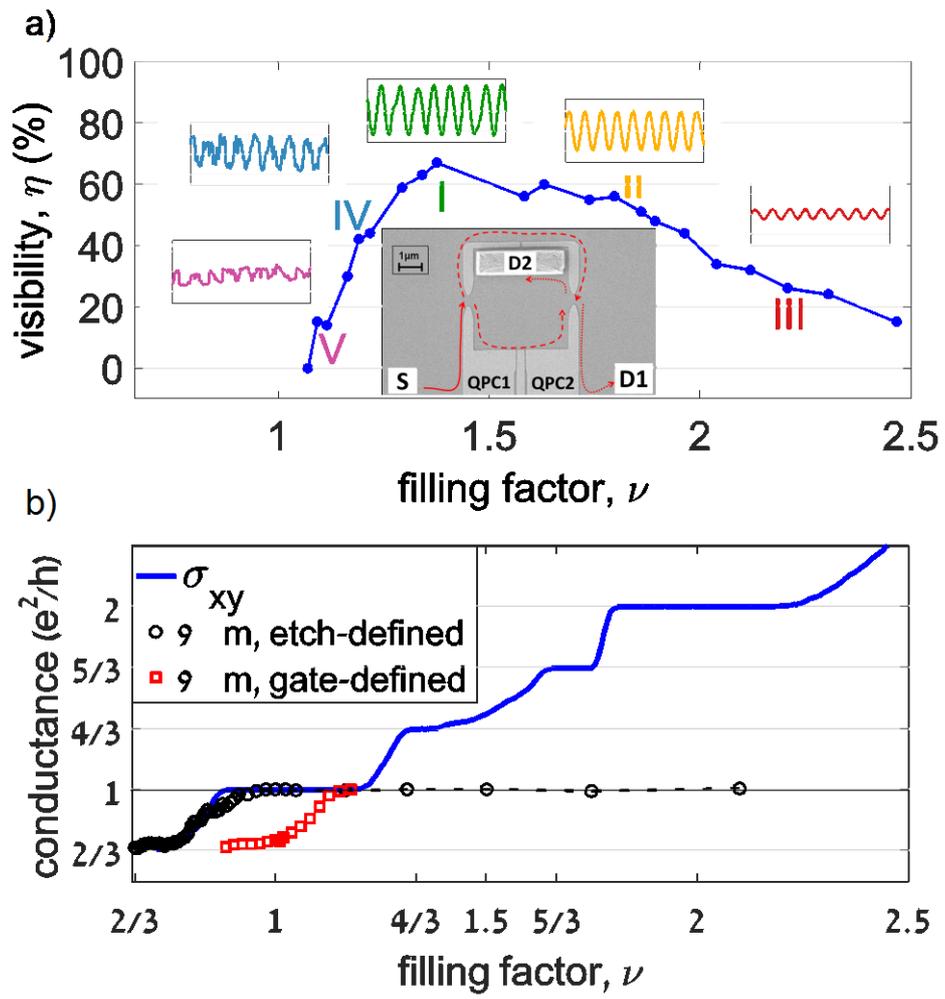

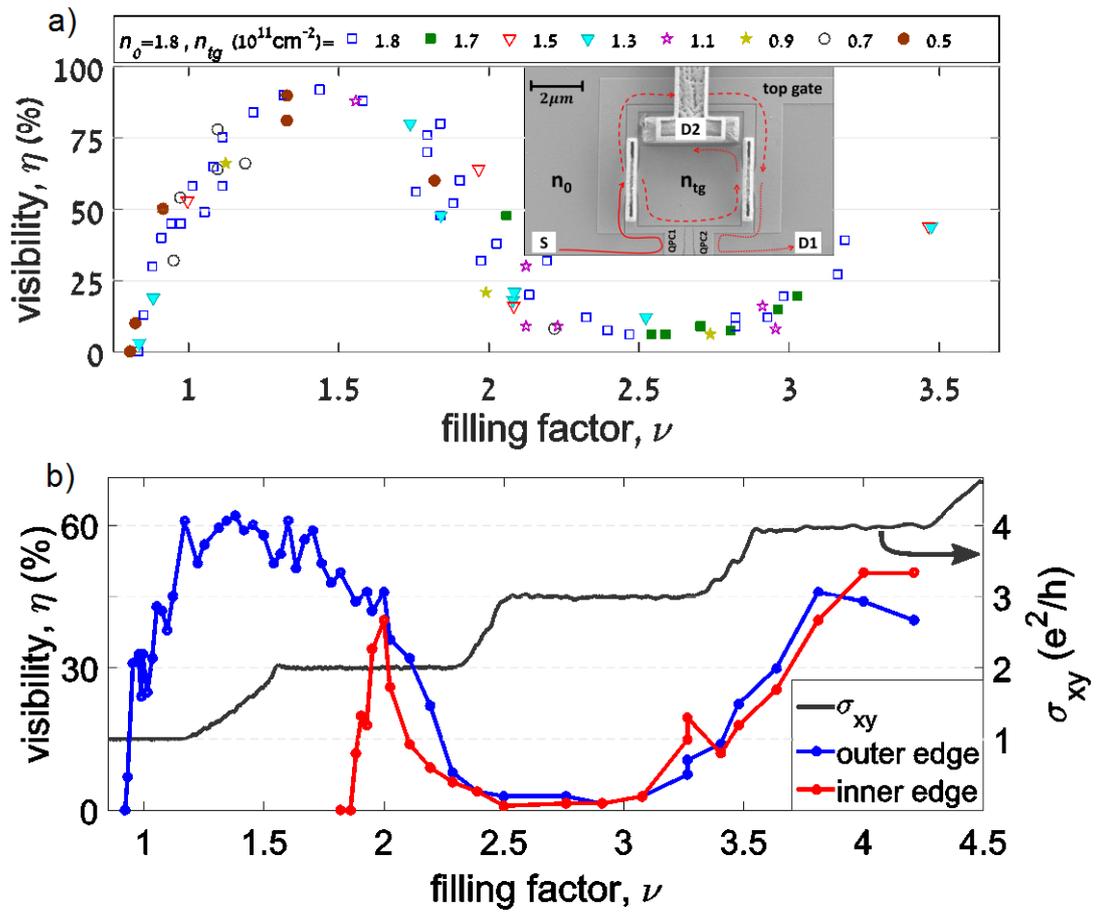

FIGURE 4

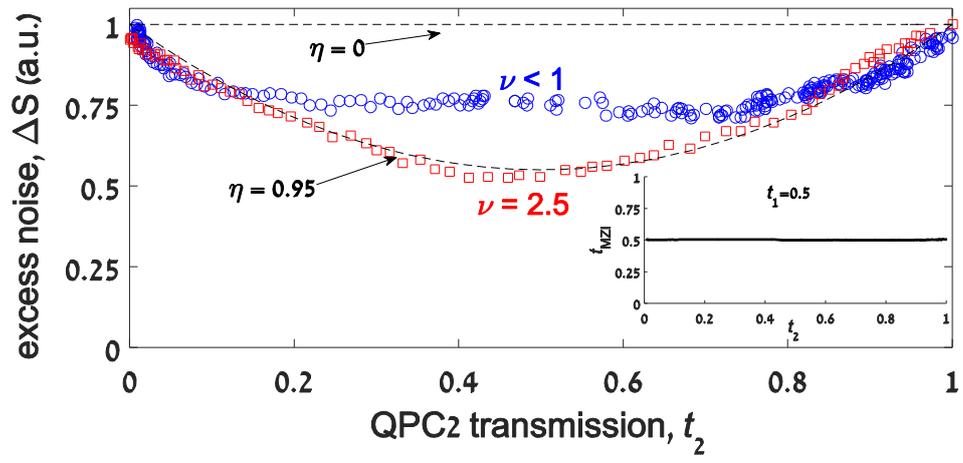

Figure Captions

Figure 1 – Visibility in a mesa defined Mach Zehnder interferometer. (a) Scanning electron microscope (SEM) image of the MZI. The paths are defined using wet etching, QPC1 & QPC2 control the transmission of each electronic beam splitter and the modulation gate (mg) can modify the lower path, such that it changes the enclosed area. The interfering channel (red arrows) can be biased either by a DC or an AC voltage, and the outcome of the MZI is either drained to the small Ohmic situated inside the interferometer (D2), or an external drain a few tens of micrometers away from it (D1). (b) Visibility of the oscillations as function of $V_{DC}$ (lobe structure). The voltage at which the visibility drops first to minimum is denoted by $V_0$, and proportional to the edge's drift velocity. Inset: an example for measured sinusoidal oscillations at D1 while scanning $V_{mg}$. (c) Visibility and velocity dependence on exact filling factor - $\nu$. For each value of $\nu$, the pre-selection QPC was set to fully transmit only the outer most channel and both the interferometer's QPCs were tuned to half transmission. The visibility was extracted by scanning $V_{mg}$ (blue circles) and velocity was calculated using the non-linear visibility (green triangles). Full and hollow shapes refer to different (yet similar) devices.

Figure 2 – Gate defined Mach Zehnder. (a) Visibility trend line as function of filling factor, shows similar behavior to the one observed for the etch-defined device. The oscillations insets along points I-V demonstrate how the oscillations diminish in different ways, whether the filling factor is higher than 1.5, showing a smooth decline, or lower, exhibiting phase jumps with increasing rate (the Y-axis of all these insets is between 90% and 10%). Main inset: SEM image of the gate defined MZI, where the metal gates control both the QPCs' transmission probability and define the two MZI's paths (the two parts of each QPC as well as D2 are connected using an air bridge, which is not shown). (b) The conductance of the outer most edge as function of filling factor, for both an etch-defined edge (black circles) and a gate-defined edge (red squares), while $\sigma_{xy}$ is at the background (blue line). It is obvious to see that only the gate-defined case supports the formation of a $\frac{2}{3}\frac{e^2}{h}$ edge despite $R_{xy}$ being at the $\frac{e^2}{h}$ plateau.

Figure 3 – (a) Visibility trend line in a gate defined MZI with a top gate (inset shows SEM picture), which is used to change the local density $n_{tg}$. It shows the visibility does not solely depend on density nor on the magnetic field, but on the ratio between the two. Each marker represents a different value of $n_{tg}$. (b) Comparison of visibility trend line between outer edge (blue) and inner edge (red). Both trends are extremely similar down to $\nu = 2$, where the inner edge visibility decays in the same fashion outer edge visibility decays around $\nu = 1$. The dark grey line is the measured $\sigma_{xy}$.

Figure 4 – Excess shot noise measurements of a MZI which shows no oscillations. $t_1$ was set to be 0.5, such that $t_{MZI} = 0.5$, the source was biased with a constant DC voltage ($50 - 80$ µV) and $t_2$ was scanned from 1 to 0, either for $\nu < 1$ (blue circles) or for $\nu = 2.5$ (red squares). In both cases $\eta$ is high (approximately 0.7 for $\nu < 1$ and 0.95 for $\nu = 2.5$), which shows that the visibility is null mainly due to phase averaging. The dashed lines are the excepted noises for $\eta = 0.95$ and $\eta = 0$. All noise data was normalized by the maximal measured noise, achieved at $t_2 = 0, 1$. Inset: an example (in this case for $\nu < 1$) how the total MZI transmission does not depend on $t_2$.